\pdfoutput=1
\documentclass{article}

\usepackage[margin = 1.5in]{geometry}

\usepackage{amsmath}
\usepackage{natbib}
\usepackage{amsfonts}
\usepackage[ruled,vlined]{algorithm2e}
\usepackage{graphicx}
\usepackage{color,soul}

\newcommand{\gjh}[1]{\textcolor{black}{#1}}
\newcommand{\strike}[1]{}
\newcommand{\zycone}[1]{\textcolor{black}{#1}}

\begin{document}

\title{\vspace{-15mm}Interpreting Models via Single Tree Approximation}

\vspace{20mm}

\date{\today}
\author{{\bf Yichen Zhou} \hspace{45mm} \footnotesize{YZ793@CORNELL.EDU} \\ 
{\bf Giles Hooker} \hspace{45mm} \footnotesize{GJH27@CORNELL.EDU} \\ 
\\
Department of Statistical Science \\
Cornell University \\
Ithaca, NY 14850, USA
}

\maketitle

\begin{abstract}
We propose a procedure to build a decision tree which approximates the performance of complex machine learning models. \gjh{\strike{We show that t}T}his single approximation tree can be used to interpret and simplify the predicting pattern of random forests \gjh{(RFs) and other models}. \gjh{The use of a tree structure is particularly relevant \strike{which are relevant structures}} in medical questionnaires \gjh{where it enables an adaptive shortening of the questionnaire, reducing response burden}. We \gjh{\strike{further}} study the asymptotic behavior of splits and introduce an improved splitting method designed to stabilize tree structure. Empirical studies \gjh{\strike{are done with} on} both simulation and real data sets \gjh{\strike{to}} illustrate that our method can simultaneously achieve high approximation power and stability.
\end{abstract}

\section{Introduction}

Random forests \citep{breiman2001random} \gjh{and other machine learning methods} have been widely used across different disciplines and are acknowledged for their outstanding predictive power \citep{caruana2006empirical}. However, machine learning models suffer from \gjh{a} trade-off between predictive accuracy and model interpretability. RFs, as \citet{breiman2001statistical} pointed out, are ``A+ predictors'' but ``on interpretability they rate F''. This paper presents a method to resolve this issue by constructing a single tree approximation that mimics the behavior of a machine learning model; \gjh{\strike{to help} both helping to} interpret \gjh{\strike{its mechanism by the predicting paths of the tree} the underlying mechanisms that produced the data, and simplifying prediction effort}. In this paper we particularly focus on the RF models, \gjh{but our procedures can be applied to mimic any prediction function}.

\subsection{Interpreting a \gjh{\strike{Random Forest} Black Box}}

\gjh{\strike{In t}T}his paper \gjh{presents an extension of \strike{we extend}} methods developed in \citet{gibbons2013computerized} where the authors have introduced the single tree approximation of a RF to produce an adaptive screening tool for depression diagnosis in psychiatry. A RF was initially built to classify the patients based on their responses to an 82-question survey, after which a decision tree was learned from pseudo samples generated by the RF to construct an adaptive screening tool with an average length of 7 questions. The aim is to reduce response burden on patients while at the same time mimicking the predictions made by the RF as much as possible. This idea is related to \citet{domingos1997knowledge} in which the author introduced Combined Multiple Models(CMM) to learn a base learner from a multiple model that is complex, incomprehensible but accurate.

Approximation trees are a useful expedient when we need to trace or reproduce the computerized decision making procedure. The predictive pattern of a tree is easy to follow and guarantees a clear predicting path which can be viewed as a guideline for reasoning. In addition, \gjh{\strike{the approximation tree uses a small subset of variables when its size is controlled} in the context of medical questionnaires, response burden can lead to poor information from long questionnaires. The tree structure requires only a subset of questions which are chosen, and presented, adaptively, considerably reducing the response effort required. \strike{This variable selection property helps in identifying variable importance, which is crucial in our psychiatry case because of question fatigue.}}

\subsection{Stablizing a Decision Tree}

Approximating an existing model by a tree is different from learning directly from data. The existing model serves as an oracle that \gjh{\strike{exploits every detail of} can be queried at any point in} the sample space from which the tree is intended to learn, thus over-fitting \gjh{ does not present a concern. \strike{is no more an issue here.}}\gjh{This means that we can make use of existing theory guaranteeing the consistency of an approximating tree. However, we also want to ensure that the resulting tree structure can be obtained reliably. The methods below are focused on ensuring that we obtain enough samples from the oracle to remove variability due to the random selection of these samples due to the construction of the tree. \strike{On one hand, asymptotic theories of decision trees have proven their consistency, which convinces us that the approximation tree will always work. On the other hand, we have to determine both the depth of the tree and how much we need to request from the existing model.}  In particular, in the context of developing an adaptive questionnaire in } \citet{gibbons2013computerized}\gjh{, our aim is to ensure that once a machine learning model is used, we reliably obtain the same approximating tree.}
\gjh{Explicitly, we adapt the strategy in }\citet{gibbons2013computerized} \gjh{of generating a large set of pseudo-examples and obtaining a prediction for each. These are then used within a CART-like algorithm to produce a tree. Here, we develop tools to ensure that enough data is available at each split so that the same tree structure is reproduced. This can be achieved via a hypothesis testing approach which allows us to control the probability that a different set of pseudo-examples would result in a different split. Importantly, large numbers of samples may be required, and these can be generated at each stage of the tree building process.}

\gjh{\strike{Since the approximation tree is built with the full access of the existing model, each split in the tree should ideally be optimal. Furthermore, as the tree serves as an interpretation of the model, we intend this interpretation to be unique. Thus the approximation tree should have a unique structure should we learn from the same model multiple times. Practically, consider the case in
Mimicking the RF by the approximation tree, different psychiatrists should obtain the same adaptive screening tool in order to achieve diagnostic consistency.}}

\gjh{\strike{Such target requests us to stabilize the approximation tree and here is our strategy. When splitting the tree, in contrast to the conventional cases of learning directly from data when we have limited sample size, we keep generating points from the existing model until we have enough confidence to conclude we will always make the same split. We will later show that by tailoring a test of better split, we can asymptotically distinguish between different splits. Since a tree is fully characterized by its splits, this strategy can guarantee the structural stability.}}

\subsection{Gini Indices}

\gjh{\strike{To quantify the split of a node we usually maximize the information gain induced by this split.} Tree building procedures select splits based on maximizing the information gain that results from each candidate split point.} There are multiple choices of defining the information gain in the literature \citep{breiman1984classification, loh1997split}. In this paper, we will focus on the Gini information associated with the Gini index as its empirical estimator. For the distribution of $(X,Y)$ where $Y \in \{1,\dots,k\}$ the category labels of $X$, one way to write the Gini gain $g$ is
\begin{align}
g  = \sum_{i\not=j}P(Y=i)P(Y=j)  = 1-\sum_{i=1}^k P(Y=i)^2.
\label{eqn:gini}
\end{align}
The empirical version of this formula defines the corresponding Gini index. It is worth noticing that this conventional definition \gjh{\strike{denotes} implies} more information with a smaller value. The formula further indicates its relation with the sample variance. In other words, smaller Gini indices implies less discrepancy among responses.

When splitting a node in the decision tree, we divide the sample space into two subsets within each \gjh{of which} the responses are more uniform \gjh{than in the whole space}, increasing the total information gain. This value is estimated by the weighted sum of the two Gini indices after splitting, hence the split with the maximal Gini index implies the best information gain and is therefore employed. In the following sections we will show that, in our approximation setting, we can determine our sample size to get more precise estimate of the Gini indices, \gjh{\strike{which} thereby} stabilizes the split at each node \gjh{\strike{and makes the tree unique}}.

\section{A test of better split}


\gjh{In this section, we will develop a means of assessing the stability of a note splitting procedure via the use of hypothesis tests. This will then be employed to ensure that we generate enough data to reliably choose the same split points. }

Consider a multi-category classification problem. The original sample consist of covariates and responses $\{(\tilde{X}_i, \tilde{Y}_i)\}_{i=1}^{n_0}$ where $\tilde{X}_i \in \mathbb{R}^m, \tilde{Y}_i \in \{1,2,\dots, k\}$, $m$ the dimension of covariate space, and $k$ the levels of responses. We obtain a RF $\mathcal{F}$ from the sample. $\mathcal{F}$ will later serve as the oracle we try to mimic, generating points (pseudo sample) $\{(X_i, Y_i)\}_{i=1}^{n}$ of arbitrary size $n$. Here $X_i=(X_i^1,\dots,X_i^m) \in \mathbb{R}^m$, and $Y_i=(Y_i^1,\dots,Y_i^k) \in \mathbb{R}^k$ are the $\mathcal{F}$-predicted class probabilities over responses. To approximate the RF performance, our tree classifiers will be constructed from $\{(X_i, Y_i)\}_{i=1}^{n}$.

\gjh{We now wish to control the probability that two different pseudo samples, $\{(X_i, Y_i)\}_{i=1}^{n}$ and $\{(X_i^*, Y_i^*)\}_{i=1}^{n}$ would result in different splits. Here, we make pairwise comparisons between the current best split, and the list of candidate alternatives. For each alternative, the p-value for a test that the difference in Gini gains is greater than zero gives us an estimate of the probability that a different data set would choose the alternative over the current best split. Summing these probabilities then gives a bound on the likelihood of splitting the current node a different way and we then select $n$ to control this probability.}


\subsection{Asymptotic Distribution of Gini Indices}

A theoretical discussion of the evaluation of splits can be found in \citet{banerjee2007confidence}. In our specific \gjh{\strike{interest}case}, we \gjh{\strike{expect to}}compare the Gini indices of \gjh{candidate} splits: \gjh{\strike{which requires to} To do so, we} examine their asymptotic behavior and obtain a central limit theorem (CLT) so normal based tests can be developed. (\ref{eqn:gini}) implies an averaging over all samples when calculating the Gini index, suggesting the existence of the CLT.

To examine two perspective splits $G_1$ and $G_2$ with the same samples, recall their Gini gains
\begin{align*}
g_1 &= 1-\pi_{1,l} \left(\sum_{j=1}^k \theta_{1,l,j}^2\right) - \pi_{1,r} \left(\sum_{j=1}^k \theta_{1,r,j}^2\right),\\
g_2 &= 1-\pi_{2,l}\left(\sum_{j=1}^k \theta_{2,l,j}^2\right) - \pi_{2,r}\left(\sum_{j=1}^k\theta_{2,r,j}^2\right),
\end{align*}
where $\pi$ represents the covariate distribution of $\tilde{X}$ and $\theta$ the conditional probability of $\hat{\tilde{Y}}$ given $\tilde{X}$. Subscripts are arranged in the order of the split, the left or right child, and the class label. For instance,
\begin{align*}
\pi_{1,l} &= P(G_1(X)=0), \\
\pi_{1,r} &= P(G_1(X)=1), \\
\theta_{1,l,j} &= P(Y=1|G_1(X)=j),
\end{align*}
and respectively for $G_2$. The empirical versions, Gini indices, are
\begin{align*}
\hat{g}_{1,n} &= 1-\frac{n_{1,l}}{n} \sum_{j=1}^k \left(\hat{\theta}_{1,l,j}\right)^2 - \frac{n_{1,r}}{n} \sum_{j=1}^k \left(\hat{\theta}_{1,r,j}\right)^2,\\
\hat{g}_{2,n} &= 1-\frac{n_{2,l}}{n} \sum_{j=1}^k \left(\hat{\theta}_{2,l,j}\right)^2 - \frac{n_{2,r}}{n} \sum_{j=1}^k \left(\hat{\theta}_{2,r,j}\right)^2.
\end{align*}

Moving to the left and right children of both splits, we denote the numbers of samples and the ratios of class labels in each child by, for $p \in \{1,2\}, j \in \{1,\dots,k\},$
\begin{align*}
n_{p,l} &= \sum_{i=1}^n 1_{\{G_p(X_i)=0\}}, \\
\hat{\theta}_{p,l,j} &= \frac{1}{n_{p,l}} \sum_{i=1}^n Y_i^j\cdot 1_{\{G_p(X_i)=0\}}, \\
n_{p,r} &= \sum_{i=1}^n 1_{\{G_p(X_i)=1\}}, \\
\hat{\theta}_{p,r,j} &= \frac{1}{n_{p,r}} \sum_{i=1}^n Y_i^j\cdot 1_{\{G_p(X_i)=1\}}.
\end{align*}
For simplicity we write, for $p \in \{1,2\},q\in\{l,r\},$
$$
N_{p,q} = \begin{bmatrix}
n_{p,q}\hat{\theta}_{p,q,1}\\
\vdots \\
n_{p, q}\hat{\theta}_{p,q,k}
\end{bmatrix}, \quad
\Theta_{p,q} = \begin{bmatrix}
\pi_{p,q} \theta_{p,q,1}\\
\vdots \\
\pi_{p,q}\theta_{p,q,k}
\end{bmatrix}.
$$
Employing a multivariate CLT we obtain
\begin{align*}
\sqrt{n} \left(\frac{1}{n}\begin{bmatrix}
N_{1,l} \\
N_{1,r} \\
N_{2,l} \\
N_{2,r}
\end{bmatrix} -
\begin{bmatrix}
\Theta_{1,l} \\
\Theta_{1,r} \\
\Theta_{2,l} \\
\Theta_{2,r}
\end{bmatrix} \right) \longrightarrow
N(0, \Sigma).
\end{align*}

To relate this limiting distribution to the difference of Gini indices we shall employ the $\delta$-method. Consider the analytic function $f: \mathbb{R}^{4k} \to \mathbb{R}$ s.t.
\begin{align*}
f(x_1, \dots, x_{4k}) = &-\frac{1}{\pi_{1,l}}\sum_{i=1}^k x_i^2
-\frac{1}{\pi_{1,r}} \sum_{i=k+1}^{2k} x_i^2 
+\frac{1}{\pi_{2,l}} \sum_{i=2k+1}^{3k} x_i^2
+\frac{1}{\pi_{2,r}} \sum_{i=3k+1}^{4k} x_i^2.
\end{align*}
The $\delta$-method imples that
\begin{align}
\sqrt{n} \left(f\left(\frac{1}{n}\begin{bmatrix}
N_{1,l} \\
N_{1,r} \\
N_{2,l} \\
N_{2,r}
\end{bmatrix}\right) - f\left(
\begin{bmatrix}
\Theta_{1,l} \\
\Theta_{1,r} \\
\Theta_{2,l} \\
\Theta_{2,r}
\end{bmatrix}\right)\right) \nonumber 
\longrightarrow N(0, \Theta^T\Sigma\Theta).
\label{eqn:main}
\end{align}

Here we write
\begin{gather*}
\Theta  = f'\left(\begin{bmatrix}
\Theta_{1,l} \\
\Theta_{1,r} \\
\Theta_{2,l} \\
\Theta_{2,r}
\end{bmatrix} \right)
= 2 \begin{bmatrix}
-\Theta_{1,l} \\
-\Theta_{1,r} \\
\Theta_{2,l} \\
\Theta_{2,r}
\end{bmatrix} \in \mathbb{R}^{4k},\\
\Sigma  = cov \begin{bmatrix}
\Theta_{1,l} \\
\Theta_{1,r} \\
\Theta_{2,l} \\
\Theta_{2,r}
\end{bmatrix} = cov \begin{bmatrix}
Y\cdot 1_{\{G_1(X)=0\}}\\
Y\cdot 1_{\{G_1(X)=1\}}\\
Y\cdot 1_{\{G_2(X)=0\}}\\
Y\cdot 1_{\{G_2(X)=1\}}
\end{bmatrix} \in \mathbb{R}^{4k \times 4k}.
\end{gather*}
We should point out that \gjh{while} (\ref{eqn:main}) provides us with the CLT we \gjh{\strike{required to qualify} need to assess} the difference between two Gini indices. After expanding (\ref{eqn:main}),
$$
\sqrt{n} \left ((\hat{g}_{1,n}-\hat{g}_{2,n}) - (g_1 - g_2)\right) \longrightarrow N(0, \Theta^T \Sigma \Theta).
$$
or asymptotically,
$$
(\hat{g}_{1,n} - \hat{g}_{2,n}) - (g_1 - g_2) \sim N \left(0, \frac{\Theta^T \Sigma \Theta}{n}\right).
$$

Hence, by replacing $\Theta, \Sigma$ by the empirical versions from the pseudo samples, we write
\begin{gather}
\hat{g}_{1,n}-\hat{g}_{2,n} \sim N\left(g_1 - g_2, \frac{\hat{\Theta}^T \hat{\Sigma}\hat{\Theta}}{n}\right).
\label{eqn:bs2}
\end{gather}

\zycone{
Denote $\Delta_n=g_{1,n}-g_{2,n}$ the Gini difference we can rewrite (\ref{eqn:bs2}) as
\begin{gather}
\hat{\Delta}_n \sim N\left(\Delta_n, \frac{\hat{\Theta}^T \hat{\Sigma}\hat{\Theta}}{n}\right).
\label{eqn:bs3}
\end{gather}
}

\subsection{Comparing Two Splits}

\zycone{
The above formula (\ref{eqn:bs2}) gives rise to the following test when comparing two splits with different batches of pseudo samples. Suppose we have two \gjh{\strike{perspective} prospective} splits $G_1$ and $G_2$\gjh{\strike{to make the split and we intend to}}. After drawing pseudo samples $\{(X_i, Y_i)\}_{i=1}^{n}$ and observing, without loss of generality, that $\hat{\Delta}_n = \hat{g}_{1,n} - \hat{g}_{2,n} < 0$. We intend to claim that $G_1$ is better than $G_2$. In order to ensure this split is chosen reliably, we can run a single-sided test to check whether we would obtain the same decision when accessing $\hat{\Delta}^*_n = \hat{g}^*_{1,n} - \hat{g}^*_{2,n} < 0$ with another independently-generated set of pseudo samples $\{(X^*_i, Y^*_i)\}_{i=1}^{n}$. Assume that $\{(X_i, Y_i)\}_{i=1}^{n}$ and $\{(X^*_i, Y^*_i)\}_{i=1}^{n}$ are independent samples, (\ref{eqn:bs3}) implies
$$
\hat{\Delta}_n^* - \hat{\Delta}_n \sim N\left(0, \frac{2\hat{\Theta}^T \hat{\Sigma}\hat{\Theta}}{n}\right),
$$
which gives,
$$
\hat{\Delta}_n^* \bigg| \left(\hat{\Delta}_n = \hat{g}_{1,n} - \hat{g}_{2,n}\right) \sim N\left( \hat{g}_{1,n} - \hat{g}_{2,n}, \frac{2\hat{\Theta}^T \hat{\Sigma}\hat{\Theta}}{n}\right).
$$
}\zycone{This distribution leads to a prediction interval based on which we would get the prediction of the Gini difference using a different pseudo sample. In order to control $P(\hat{\Delta}_n^*<0)$ at a confidence level $1-\alpha$, we need
\begin{eqnarray}
\label{eqn:bs}
\hat{g}_{1,n} - \hat{g}_{2,n} < Z_{\alpha}\cdot \sqrt{\frac{2\hat{\Theta}^T \hat{\Sigma}\hat{\Theta}}{n}},
\end{eqnarray}
}where $Z$ is the inverse c.d.f. of a standard normal. With a sufficiently large $n$ it is possible to always determine the better split between $G_1$ and $G_2$ should they have any difference. In addition, by combining this test with \gjh{a} pairwise comparisons \gjh{procedure}, we are capable of finding the best split among multiple prospective splits.

\subsection{Sequential Testing}

The power of \gjh{\strike{the} this} better split test increases with $n$. Since we need to determine $n$ to reveal any detectable difference between two splits, when no prior knowledge is given regarding the magnitude of the difference, we need an adaptive approach to increasing $n$ accordingly.

\gjh{\strike{Provided} For a fixed} confidence level $\alpha$, suppose we have tested at sample size $n$ and get p-value $p_n > \alpha$. Referring to (\ref{eqn:bs}), we have
$$
\sqrt{n}\cdot \frac{\hat{g}_{1,n}-\hat{g}_{2,n}}{\sqrt{2\hat{\Theta}^T \hat{\Sigma} \hat{\Theta}}} = Z_{p_n}.
$$
Notice that $\dfrac{\hat{g}_{1,n}-\hat{g}_{2,n}}{\sqrt{2\hat{\Theta}^T \hat{\Sigma} \hat{\Theta}}}$ is the estimator of $\dfrac{g_1-g_2}{\sqrt{2\Theta^T \Sigma \Theta^T}}$ which is an intrinsic constant with respect to the pairwise comparison. Hence in order to reach a p-value less than $\alpha$ we may increase sample size to $n'$ such that
\begin{align*}
\sqrt{n'}\cdot \frac{\hat{g}_{1,n}-\hat{g}_{2,n}}{\sqrt{2\hat{\Theta}^T \hat{\Sigma} \hat{\Theta}}} = Z_{\alpha},
\end{align*}
which yields that
\begin{gather}
\sqrt{\frac{n}{n'}} = \frac{Z_{p_n}}{Z_{\alpha}}.
\label{eqn:ad1}
\end{gather}

Due to pseudo sample randomness, a few successive increments are required before we land in the confidence level. We also need an upper bound for $n'$ \gjh{and a default split order} in case the difference between two splits is too \gjh{\strike{insignificant} small} to identify.

\subsection{Multiple Testing}
So far we have obtained a method to compare a pair of splits. When splitting a certain node, however, it is always the case that we need to choose the best split among multiple $G_1,\dots G_m$. In order to adapt our pairwise better split test to this situation, we consider modifying the problem slightly into deciding whether the split with the minimal Gini index is intrinsically superior than any other splits. This problem can be resolved by conducting pairwise comparisons of the split with minimal Gini index against the rest.

If we still want to test at a certain significance $\alpha$ \gjh{\strike{that}} whether the \gjh{\strike{one} split} with the lowest estimated Gini index, i.e, $\hat{g}_{n,(1)}$, is the optimal, we can still work within the scheme of the pairwise comparison with an additional procedure controlling the familywise error rate (FWER). Here we make an analogue of the Bonferroni correction \citep{dunnett1955multiple}.

$\bullet$ Test the hypotheses $H_{i,0}: g_{(1)} = g_{(i)}, i=2,\dots, t$. Get the $p$-values $p_2,\dots,p_t$.

$\bullet$ Analogous to a Bonferroni correction, use $\sum_{i=2}^t p_t$, the upper bound for making at most one Type I error, as the $p$-value of the multiple comparison.

This test aggregates all significance levels into one. The Bonferroni correction will result in a \gjh{\strike{more}} conservative \gjh{\strike{estimation} estimate} as we ignore much \gjh{of the} correlation structure of the splits. In this scenario, the updates of sample size made in sequential testing should also be adapted as we are now taking the aggregated significance level. A quick and feasible fix is to replace the $p_n$ in (\ref{eqn:ad1}) by the aggregated significance level. Alternatively, we may just test between the best two splits.

Because of the computational cost, when we have two splits that \gjh{\strike{too similar to distinguish} cannot be distinguished}, the sequential and multiple testing \gjh{procedure} may end up demanding an extremely large number of points to make the test significant. In practice, we halt the testing early at a cutoff of certain amount $N_{ps}$ of points, and choose the current best split. This compensation for computation time might lower the real power of the test, leading to a less stable result.

\subsection{Choice of Prospective Splits}
\gjh{\strike{We use a naive method to choose prospective splits at each node.}
In building an approximation tree, we only consider making splits at those points which would have been employed in a tree generated from the original training data. \strike{Again w}W}e look at the original samples that have been carried along the path and take the possible combinations of the covariates and their middle points of adjacent values that have appeared in those samples.

Although this method will \gjh{\strike{at the beginning} initially} generate a large number of prospective splits, because of the sequential testing scheme, most of those splits will be identified \gjh{as} far worse than the best after a few tests and can be discarded, leaving a negligible effect on the overall performance. In practice, we implement a scheme \citep{benjamini1995controlling} to adaptively discard splits that perform far worse than the current best. All splits are ordered by their p-values against the current best split, and the splits fall below the threshold are discarded.

\section{Approximate a RF}

To build an approximation tree, we replace the greedy splitting criterion by our stabilized version within the CART construction algorithm. At each node, we generate an initial \gjh{\strike{amount} number} of sample points from the RF, guaranteeing that those points belong to this node. Then we compare prospective splits simultaneously based on this set and decide whether we choose the one with the smallest Gini index with certain confidence or request more pseudo sample points. In the latter case, we keep generating until the pseudo sample size reaches what \gjh{is} required by the sequential testing \gjh{procedure}. This is repeated until we distinguish the best split. We perform this procedure on any node that needs to split during construction to get the final approximation tree.

\begin{algorithm}
\DontPrintSemicolon
\KwData{Random forest $\mathcal{F}$, covariate distribution of $X$}
\KwResult{Approximation tree $\mathcal{T}$}
\SetKwFunction{SplitNode}{SplitNode}
{\bf Function} \SplitNode{node V}{

    \eIf{V satisfies stopping condition}{
        stop and return
    }{
        generate $n$ pseudo samples from $\mathcal{F}$ \;
        find prospective splits $G_1,\cdots, G_m$.\;
        \While{cannot distinguish the best split among $G_1,\cdots, G_m$}{
            generating more pseudo samples whose size is decide by the sequential testing
        }
        claim the best split among $G_1,\cdots, G_m$ and split V by it\;
        \SplitNode{V's left child} \;
        \SplitNode{V's right child}\;
    }
}
$\mathcal{T} \leftarrow$ \SplitNode{root} \;
\caption{Approxmation Tree}
\end{algorithm}

There are several parameters to tune for this algorithm. We first need all the parameters for CART construction, i.e., the maximal depth of the tree, or maximal and minimal number samples in each leaf node. \gjh{\strike{Besides, we have} We must also choose} $\alpha$ \gjh{\strike{which controls} to control} the significance of the test of better split, and $N_{ps}$ which controls the maximal amount of psuedo samples we require at each node.

\subsection{Generating Points \gjh{\strike{from a Random Forest}}}
To generate the pseudo sample, we first generate pseudo covariates then \gjh{obtain predictions from the RF \strike{employ the given random forest on them}}to get the responses. It is worth noticing that the first step here may encounter the obstacle that, in practice, we do not have the prior distribution of covariates.

Conventional statistical techniques apply here. Some methods focus on estimating the underlying distribution by smoothers \citep{wand1994kernel}, while the others, for instance, bootstrapping or residual permutation, attempt to directly manipulate and reorganize the samples to generate more samples. In the purpose of exploring more of the covariate space, we\gjh{\strike{would like to}} take the first approach and use a Gaussian kernel smoother upon the empirical distribution of the samples. This translates to generating pseudo covariates from observed covariates plus random noise. \gjh{In the case of discrete covariates, we choose a neighbouring category with a small probability.}

When we go further down the approximation tree, the covariate space may as well be narrowed down by the splits along the path. A feasible covariate generator can thus be produced by only smoothing the empirical distribution of those original samples that have been carried on by this path. We further check the boundary condition to ensure that the covariates we generated agree within the region divided by the splits along the path.

\section{Empirical Study}

We have conducted \gjh{\strike{two}}empirical studies on both simulation and real data to show \gjh{how} the performance \gjh{ of \strike{when our}} approximation tree compares with both decision trees and the \gjh{original} RF. The performance is \gjh{mainly} assessed in two \gjh{\strike{aspects} ways}:\gjh{\strike{the}} consistency with the RF, and\gjh{\strike{the}} stability.

In order to evaluate consistency, we generate new covariates and measure how much the predictions of approximation tree agree with \gjh{\strike{the ones} those} of the RF. In this paper we are more interested in the mimicking ability rather than the predictive power of the approximation tree. However, we will still compare the \gjh{predictive accuracy} of approximation trees with decision trees \gjh{\strike{on the predictive accuracy}}.

To measure\gjh{\strike{the}} stability, which is defined in our case as the structural uniqueness, we construct multiple approximation trees out of a single RF and look into the variation in their structures. The better split test does not always guarantee a consistent pick through multiple trials due to the pseudo sample randomness, hence we hope to see small variation among all the trees built. We also examine the trees at different depths to capture the variation along the tree growth.

\subsection{Simulated Data}

We assume that the original samples have $\tilde{X} \in \mathbb{R}^5$ and $\tilde{Y} \in \{0, 1\}$. The covariate distribution is given by $\tilde{X} = (x_1,\dots,x_5) \sim \mbox{Unif}[0,1]^5,$ and $\mbox{logit}(P(\tilde{Y}=1|\tilde{X}))$ is given by
\begin{gather*}
\begin{cases}
2, &x_1>0.5, \quad x_2>0.7,\\
-3, &x_1>0.5,  \quad 0.7 \geq  x_2>0.2,\\
-4, &x_1>0.5,  \quad x_2 \leq 0.2, \\
3, &x_1 \leq 0.5,  \quad x_5 \leq 0.5, \quad x_3+x_4^2 \geq 1.4, \\
2, &x_1 \leq 0.5,  \quad x_5 \leq 0.5, \quad 1.4 > x_3+x_4^2 \geq 0.5, \\
-2,  &x_1 \leq 0.5,  \quad x_5 \leq 0.5, \quad x_3+x_4^2 < 0.5, \\
2, &x_1 \leq 0.5,  \quad x_5 > 0.5.
\end{cases}
\end{gather*}

\subsubsection{Predictive and Mimicking Accuracy}

We compare across three methods: classification trees (CART), random forests (RF) and our proposed single tree approximation (STA). We generate 1,000 sample points from this distribution and \gjh{\strike{employ} obtain} a standard RF consisting of 100 trees and a \gjh{single} classification tree \gjh{\strike{to fit} from them}. Then we build our approximation tree via the algorithm\gjh{\strike{mentioned}} above. The significant level $\alpha$ for the test of better split is set to be 0.1, and the maximal number of pseudo samples at each node $N_{ps}$ is set to be 10,000, 100,000, and 1,000,000 respectively. Both CART and STA set the maximal tree depth to be 5 including the root.

\begin{figure}[htbp]
    \centering
    \includegraphics[width=4.5in]{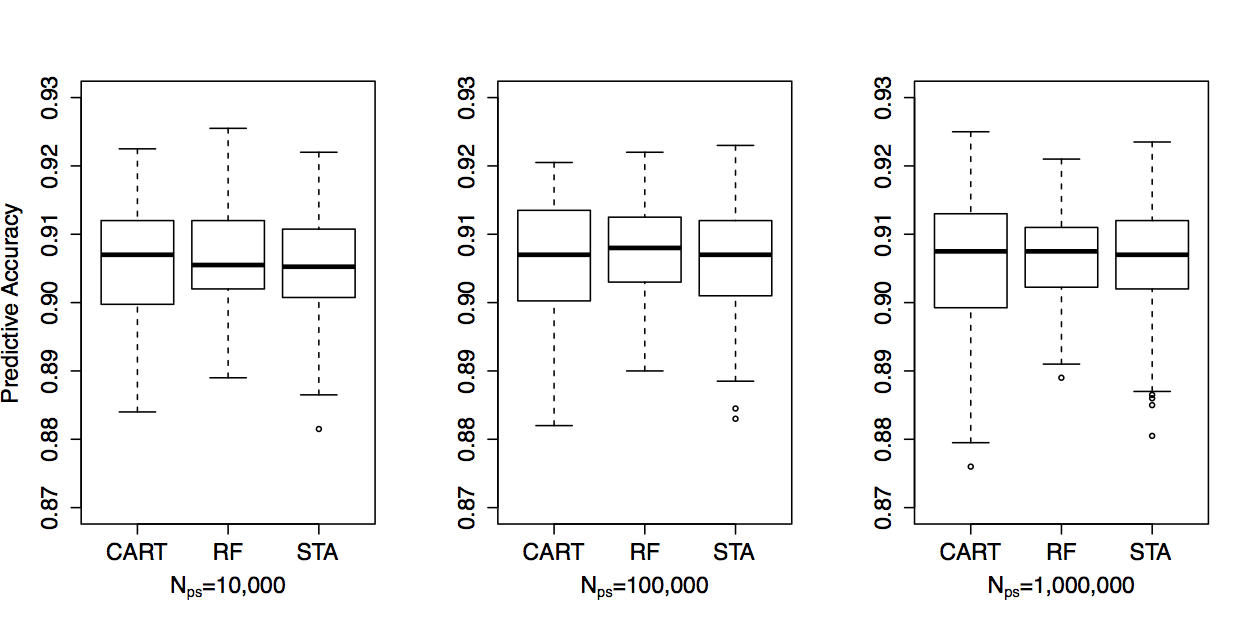}
    \caption{Predictive Accuracy. Three methods here are classification tree (CART), random forest (RF) and single tree approximation (STA) respectively.}
    \label{fig:PA1}
\end{figure}

Figure \ref{fig:PA1} shows the predictive accuracy of the three methods on new test points. On average they share similar predictive accuracy\gjh{\strike{,}; \strike{whereas}} the RF has the smallest variance, followed by the single tree approximation. This coincides with our expectation that STA is capable of inheriting stability from the random forest. Since the relation between the covariates and the responses is relatively simple, the increase of $N_ps$ has not introduced a significant improvement in performance.

\begin{figure}[htbp]
    \centering
    \includegraphics[width=4.5in]{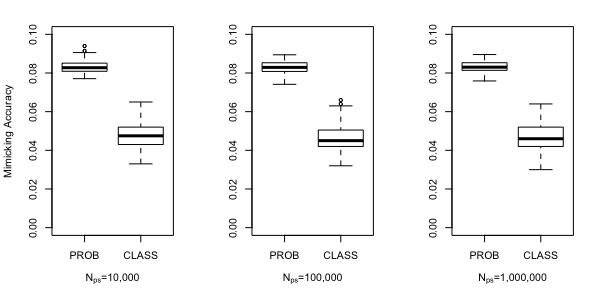}
    \caption{Mimicking Accuracy. PROB compares the output of RF and STA by the $L^1$ difference of their class probabilities. CLASS compares the output by the predicted class labels.}
    \label{fig:MA1}
\end{figure}

Figure \ref{fig:MA1} shows the comparison between RF and STA in terms of the $L^1$ difference of their predicted class probability, and the disagreement of their class labels. Again the increase of $N_{ps}$ has no significant improvement in performance. The approximation trees, which are solely built to their fifth layer, already have achieved \zycone{around 95\%} agreement on average with the RFs. It is reasonable to believe that by expanding the trees to larger sizes the mimicking accuracy can still be marginally increased by ``overfitting'' the RFs.

\subsubsection{Stability}

\begin{figure}[htbp]
    \centering
    \includegraphics[width=4.5in]{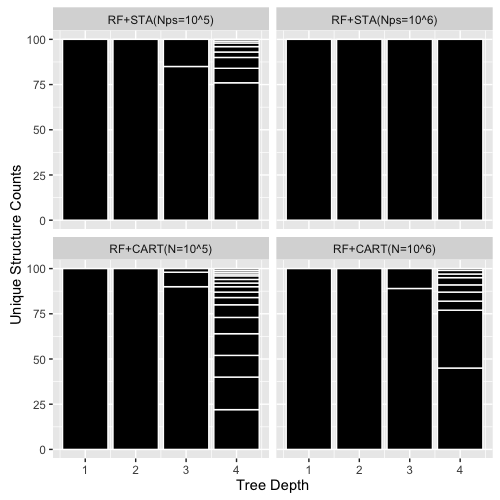}
    \caption{STA stability. Four settings: RF+STA with $N_{ps}=100,000$, RF+STA with $N_{ps}=1,000,000$, RF+CART with $N=100,000$ sample points, RF+CART with $N=1,000,000$ sample points. In each plot, the columns represent the depths of the trees (from 1 to 4 excluding the root). In each column, a single black bar represents a unique structure of the tree, while the height of the bar represents the number of occurrence of that structure out of 100 replications.}
    \label{fig:syntree1}
\end{figure}

\begin{figure}[htbp]
    \centering
    \includegraphics[width=4.5in]{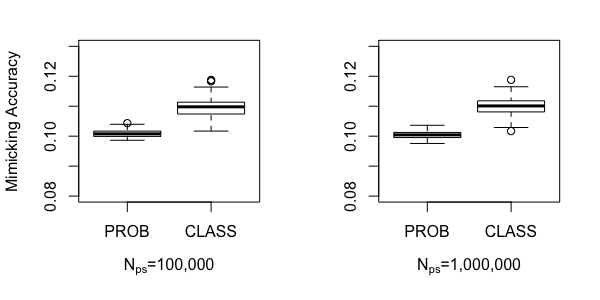}
    \caption{Mimicking Accuracy. PROB compares the output of RF and STA by the $L^1$ difference of their class probabilities. CLASS compares the out by the predicted class label. From left to right represent $N_{ps}=100,000$ and $N_{ps}=1,000,000$.}
    \label{fig:mddacc}
\end{figure}

We compare the stability of STA in contrast to CART under \gjh{a} similar \gjh{\strike{construction}} setting. Following the same setting as the previous experiment, we first employ a RF to learn the sample points, then draw pseudo sample points from the RF to generate both STA and CART. In terms of CART, to obtain a fair comparison, \gjh{\strike{each time}} we generate $N_{ps}$ pseudo samples to build a complete CART tree\gjh{\strike{. $N$ is controlled at the same magnitude of $N_{ps}$}}, the maximal number of pseudo sample points STA can utilize\gjh{.\strike{, so}} CART will \gjh{\strike{take} therefore have a} slight advantages when generating the top layer splits.

Figure \ref{fig:syntree1} shows the result \gjh{\strike{as} of} the two methods labeled as RF+STA and RF+CART. Defining \gjh{\strike{the}} stability as \gjh{\strike{the}} structural uniqueness, we examine each split by both the \gjh{chosen} covariate and the value \gjh{used to split}. CART has very poor stability when $N$ is small \gjh{and} when the trees are \gjh{\strike{viewed till the fourth layer splits} continued to four layers}. Our STA, in contrast, has much fewer structures. The variation of the approximation trees shrinks along with the increase of $N_{ps}$. \zycone{When $N_{ps}$ is $1,000,000$, STA still has a unique instances at the forth layer splits.} \zycone{The results meet our expectation that STA achieves additional structural stability with the test we proposed during CART tree construction.}

\subsection{Major Depression Disorder (MDD) Data}
We now demonstrate our method using the MDD data set which motivated \citet{gibbons2013computerized}. In this study, we focus on examining the stability of approximation trees and the consistency between the trees and the RF. The data set consists of 836 patients and 88 integer-valued covariates representing their responses to 88 survey questions. Among 88 questions there are 27 yes-no questions, 11 four-choose-one's and 50 five-choose-one's. Patients are \gjh{\strike{binarily}} classified as either with severe depression (257 out of 836) or without. \gjh{\strike{Again, w}W}e obtain a RF from these samples, and then mimic the RF with 100 approximation trees (STA). We set $N_{ps}$ to be 100,000 and 1,000,000 respectively. The significant level $\alpha$ for the test of better split is set to be 0.1.

Figure \ref{fig:mddacc} shows the mimicking power of the approximation trees. Notice that by constructing the trees to the fifth layer, the approximation trees already agree on about 90\% of the classification results\gjh{. \strike{, as well as a 0.1} The} $L^1$-norm difference in \gjh{\strike{terms of the}} class probabilities \gjh{is 0.1}. Note that by increasing $N_{ps}$ the accuracy does not significantly increase, which demonstrates that the decision tree prediction performs well \gjh{\strike{regardless of the} at a reasonable} number of pseudo samples.

\begin{figure}[htbp]
    \centering
    \includegraphics[width=4.5in]{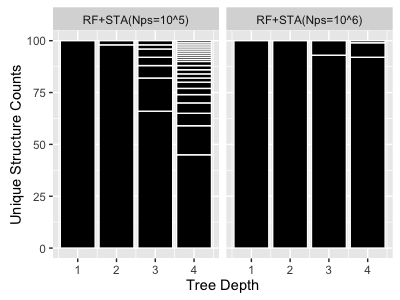}
    \caption{STA stability. STA with $N_{ps}=1,000,000$ reports 10 structures when the trees are build to the fifth layer, with the major structure occurring 68 out of 100 times.}
    \label{fig:mdd0}
\end{figure}

Figure \ref{fig:mdd0} \gjh{\strike{shows} demonstrates} the stability \gjh{ of our procedure}. The stability increases along with the increasing $N_{ps}$, the maximal amount of pseudo samples generated for deciding splits. In addition, the case $N_{ps}=1,000,000$ brought up several \gjh{\strike{insightful facts} relevant observations}. Our $\alpha$, the significance level of the test of better split, works conservatively here as all trees have the same root and second-layer splits. \zycone{Also, 92 out of the 100 trees have the exactly same structure of their top four-layer splits that represents an identical 5-layer tree.} This disagreement ratio among all the 100 trees we built is decently low\gjh{\strike{, therefore by f}}. Following \citet{gibbons2013computerized} we improve the depression diagnosis from an 88-question survey plus a RF predictor,  to \gjh{obtain} a unique adaptive screening tool with at most 4 questions, while retaining 90\% consistency. \gjh{\strike{Furthermore, we can still build the tree deeper to achieve better fit of the random forest.}}

\gjh{\strike{We also get two observations here. First, though} We also note that although} we only have 836 original patients, we \gjh{\strike{might} may} still need over 1,000,000 pseudo points to \gjh{\strike{fix} stabilize} a split. One reason to this \gjh{\strike{issue}} is that we have \gjh{\strike{too}} many variables and values to choose as a splitting rule, which is likely to cause the best splits \gjh{ to be} indistinguishable \gjh{\strike{. Surely this requires} requiring} a stopping rule \gjh{ for $n$}. \gjh{\strike{Under relatively high dimensional setting this issue is intractable, whereas a possible} An alternative} remedy is to obtain a prior set of fewer splits of interests and split by this set. However, in general we still require a large amount of points.

Another observation is that the significance level $\alpha$ controls the stability at a split-wise level. It is possible extend this to further stabilize the tree by again introducing the FWER at the tree level. Notice this procedure may also increase the number of pseudo samples we need at each split.

%
\section{Conclusion}

%

In this paper we have proposed a procedure to mimic a complex model \gjh{\strike{, especially a random forest,}} by a single decision tree approximation. The asymptotic behavior of differences of Gini indices was studied and a normal based test of better split was developed and inspected under this oracle learning setting. We further incorporated this test in \gjh{a} tree building \gjh{procedure} to ensure the performance and the stability of the approximation trees. Empirical studies were done based on both simulation and real data and the results met our expectation. We in addition presented the interpretability of our procedure on the real data.

\gjh{While we have used RFs as a oracle learning method, our algorithm can be applied with any machine learning tool.  These methods can be readily extended to alternative splitting criteria and to regression problems.  Important future problems include methods to choose the optimal depth of an approximation tree. This could be based simply on predictive performance. Alternatively, where distributional results are available for predictions from the original learner as in \citet{mentch2014ensemble}, a hypothesis tests of the difference between the original and approximated predictions could also be employed.}

\section*{Acknowledgements}
This work was partially supported by grants NIH R03DA036683, NSF DMS-1053252 and NSF DEB-1353039.  The authors would like to thank Robert Gibbons for providing the CAT-MDD data.

\bibliography{citation}
\bibliographystyle{plainnat}
\end{document}